\documentclass[12pt,preprint,natbib]{aastex}
\begin{document}
\title{A Deep Keck/NIRC2 Search for Thermal Emission from 
 Planetary Companions Orbiting Fomalhaut}
\author{Thayne Currie\altaffilmark{1}, 
Ryan Cloutier\altaffilmark{1},
John H. Debes\altaffilmark{2},
Scott J. Kenyon\altaffilmark{3},
%Timothy J. Rodigas\altaffilmark{3,4},
Denise Kaisler\altaffilmark{4}
}
%,
%Adam Burrows\altaffilmark{6}, Ryan Cloutier\altaffilmark{1}}
%, 
%Ray Jayawardhana\altaffilmark{1},
%Scott J. Kenyon\altaffilmark{4}}
\altaffiltext{1}{University of Toronto}
\altaffiltext{2}{Space Telescope Science Institute}
\altaffiltext{3}{Harvard-Smithsonian Center for Astrophysics}
%\altaffiltext{3}{University of Arizona}
%\altaffiltext{4}{Department of Terrestrial Magnetism, Carnegie Institution of Washington}
\altaffiltext{4}{Citrus College}
%\altaffiltext{6}{Department of Astrophysical Sciences, Princeton University}
\begin{abstract}
We present deep Keck/NIRC2 1.6 and 3.8 $\mu m$ imaging of Fomalhaut 
to constrain the near-infrared (IR) brightness of Fomalhaut \lowercase{b}, 
recently confirmed as a likely planet (Currie et al. 2012a), 
and search for additional planets at $r_{proj}$ = 15--150 $AU$.  Using advanced/novel PSF subtraction 
techniques, we identify seven candidate substellar companions 
 Fomalhaut b-like projected separations.  However,
 multi-epoch data shows them to be background objects.  
%One of these is likely the candidate oint source identified from Spitzer/IRAC data from Janson et al. (2012).
We set a new 3-$\sigma$ upper limit for Fomalhaut b's $H$-band brightness of m($H$) $\sim$ 23.15
 or 1.5--4.5 $M_{J}$.
%, depending on the system age and planet cooling models used.  
We do not recover the possible point source reported from Spitzer/IRAC data:
 at its location detection limits are similar to those for Fomalhaut b.  
Our data when combined with other recent work rule out planets with masses 
and projected separations comparable to HR 8799 bcde and $M$ $>$ 3 $M_{J}$ planets
at $r_{proj}$ $>$ 45 $AU$.  
  %Thus, a luminous debris disk, even one with dynamical signatures, is not always 
%a signpost for super Jovian-mass companions, although Fomalhaut may not be atypical 
%as our mass limits are comparable to those obtained for many, more distant debris disk-bearing stars.  
The \textit{James Webb Space Telescope} will likely be required to shed substantial further light on Fomalhaut's planetary 
system in the next decade.
\end{abstract}
\section{Introduction}
The nearby A star Fomalhaut has long been suspected of harboring a planetary system, given 
its bright, dusty Kuiper belt-like debris ring \citep{Aumann1985,Stapelfeldt2004,Acke2012} 
whose pericenter offset from the star is consistent with dynamical sculpting from
 an embedded planet \citep{Kalas2005}.  
 Using the \textit{Hubble Space Telescope}'s \textit{Advanced Camera for Surveys} (ACS), 
\citet{Kalas2008} identified a candidate companion, Fomalhaut b, thought to 
be consistent with a jovian planet sculpting the debris ring \citep{Chiang2009} whose 
emission is at least partially due to a planet atmosphere and 
circumplanetary accretion.  However, sensitive IR observations 
failed to recover Fomalhaut b \citep{Marengo2009,Janson2012}, raising doubts about 
its status as a planet and whether the claimed detection was spurious.

However, in late 2012 \citet{Currie2012a} announced the confirmation of Fomalhaut b, recovering 
it at a high signal-to-noise in multi-epoch data in filters where \citealt{Kalas2008} reported detections -- F606W 
and F814W -- and reporting a new ACS detection at F435W.  With detections at three filters and new 
IR upper limits at $J$ band (1.25 $\mu m$), \citet{Currie2012a} showed that Fomalhaut b's emission 
can be completely explained by scattered starlight from small dust, not thermal emission or accretion,
but that this dust is likely bound to a planet-mass body.
%, a conclusion supported by 
%more in-depth analyses (S. Kenyon et al. 2013, in prep.).
Soon thereafter, \citet{Galicher2013} independently
reported their recovery of Fomalhaut b at the three ACS filters, presented an optical detection from 2010 STIS 
data\footnote{\citet{Currie2012a} also reported a detection of Fomalhaut b in the same STIS data but 
did not formally present this as a result.}, and 
provided additional, sensitive near-IR upper limits using \textit{WFC3}.

While Fomalhaut b is a real object and likely identifies a planet-mass companion, 
the system may harbor additional planets.
In addition to a cold debris ring, Fomalhaut has a warm, asteroid belt analogue located far closer to the star 
\citep[$\approx$ 10--20 $AU$;][]{Su2013,Stapelfeldt2004}.  This configuration is similar to that for another A star, HR 8799 \citep{Su2009}, 
which has debris populations at $\approx$ 9 $AU$ and $\approx$ 95 $AU$ with 
 four directly-imaged planets located at $a_{proj}$ $\approx$ 15--70 $AU$ 
between the debris populations \citep{Marois2011,Currie2011a}.  
Preliminary results on Fomalhaut b's astrometry may point to a high eccentricity, 
ring-crossing orbit due to dynamical perturbations from another planet
\citep[][though see Galicher et al. 2013]{Kalas2013}.
% (though see Galicher et al. 2013).
Finally, it is unclear whether Fomalhaut b -- or another object -- is sculpting 
the debris ring: \citet{Janson2012} identify a candidate point source located interior to Fomalhaut's debris ring 
but on the opposite side whose brightness would be consistent with that from a jovian mass companion. 

In this Letter, we present new ground-based high-contrast imaging limits on the infrared brightness of 
Fomalhaut b
 and a search for the unseen planets responsible for sculpting the host star's debris belts 
from 2\farcs{}0 to 20\farcs{}0 ($\approx$ 15--150 AU).  This work complements recent searches 
conducted over narrower/smaller ranges in angular separations \citep{Kenworthy2009,Kenworthy2013},
presenting the first deep constraints on a hypothetical Fomalhaut `c' at $\approx$ 4--6\arcsec{} or a 
projected separation of $\approx$ 30--45 $AU$, similar to the inner two HR 8799 planets $c$ and $d$ 
\citep{Marois2008}.
%, and sampled over all position angles.

\section{Data}
\textbf{Observations} -- We downloaded 2002--2005 NIRC2 data from the Keck Observatory Archive (KOA)
 in the $H$ (1.65 $\mu m$) and $L^\prime$ (3.8 $\mu m$) filters 
 (PIs B. Zuckerman, E. Becklin, and P. Kalas; Table 1) 
obtained either as a part of the first large ground-based exoplanet imaging surveys \citep[e.g.][]{Kaisler2005} 
or as targeted observations of Fomalhaut after the discovery of its debris ring \citep{Kalas2005}.
  The 2002-2003 Fomalhaut data sets were obtained in classical imaging; others were obtained in 
\textit{angular differential imaging} \citep[ADI][]{Marois2006}.  
\citet{Kalas2008} previously report $H$-band upper limits for Fomalhaut b from Keck/NIRC2 2005 
data.  
%Detections/upper limits at other separations from these data and any results 
%from the other data have not been presented in refereed literature.
The data include those obtained in the wide camera (39.69 mas/pixel) able to identify objects 
at wide separations like Fomalhaut b and the narrow camera \citep[9.952 mas/pixel;][]{Yelda2010}, 
sensitive to objects within $\sim$ 5\arcsec{}.  Basic image processing steps followed previous
 NIRC2 reductions \citep{Currie2012b, Currie2012c}.  We adopt the narrow camera distortion 
correction used in \citet{Yelda2010} and the empirically-determined wide camera correction 
(kindly provided by Hai Fu).

\textbf{PSF Subtraction} -- Although the 2002--2003 data were obtained in classical imaging, we nevertheless 
were able to apply PSF subtraction techniques normally restricted to ADI-mode data (3rd paragraph).  
Briefly, for alt-az telescopes like Keck, in ADI mode objects located off-axis from a star change in position 
angle on the detector while quasi-static speckles remain fixed.  In classical imaging 
such objects stay fixed in position while much of the speckle pattern rotates with time.  However, the rotation rate of 
the speckle pattern, like for off-axis point sources in ADI data, is equal to the rate-change of the parallactic 
angle.  

Thus, we can mimick an ADI dataset with frames $i$ = 1...$N_{tot}$ obtained in classical imaging by \textit{rotating} 
each image $i$ by its change in parallactic angle from the first image: $\delta$$\theta_{i}$ = -1$\times$($PA_{o}$-$PA_{i}$).
To realign each image with north-up after PSF subtraction, we \textit{derotate} each image $i$ by the same $\delta$$\theta_{i}$,
as in a normal ADI data set.  Applying this method, we partially suppress quasi-static speckles and 
achieve significantly deeper contrast gains than simple unsharp masking, subtraction using a 180-degree rotated PSF, 
or subtraction with a PSF reference star.  This method has been independently and successfully tested, 
yielding a detection of HR 8799 c from July 2004 NIRC2 classical imaging data presented in \citet{Marois2008} (C. Marois, 2013 
pvt. comm.) and can be applied to numerous KOA classical imaging data to identify new planets.

For PSF subtraction, we explored several approaches.  For the $H$-band data and regions beyond 
3\arcsec{} in the $L^\prime$ data we used $A-LOCI$ with default values of a large rotation gap ($\delta$ = 1.5--3), 
a large optimization area ($N_{A}$=300--500) and a weak \textit{singular value decomposition} ($SVD$) 
cutoff of 10$^{-6}$ \citep[see][for definitions]{Lafreniere2007,Marois2010b,Currie2013}.  
For the inner 3\arcsec{} of the $L^\prime$ data, we achieved a slight further gain using smaller optimization areas, 
while adding pixel masking and speckle filtering \citep{Currie2012a} and retaining the same $SVD$ cutoff. 

\textbf{Photometric Calibration} -- To flux-calibrate the $H$ and $L^\prime$ data and derive contrast limits, we calculate
the brightness of Fomalhaut A observed through the mask in an aperture equal to the measured FWHM and correct for the pupil mask's extinction.  
For the $H$-band data, we adopt the extinction estimate from \citet{Metchev2004} of 7.79 $\pm$ 0.22 mags.  
We derive the $L^\prime$ coronagraph extinction from comparing the brightness of Fomalhaut as observed through the mask to the A-type star 
SAO 183818, which was observed immediately after Fomalhaut on July 28, 2003 (m(L$^\prime$) $\sim$ 7.2 $\pm$ 0.1): 
$m_{ext} (L^\prime)$ $\sim$ 6.06 $\pm$ 0.15.  
%, where uncertainties include the sky background, brightness variations in SAO 183818 data, and fluctuations 
%in Fomalhaut's brightness from behind the coronagraph.  
%For the 2009 $L^\prime$ data, we use unsaturated images of Fomalhaut for
%flux calibration.  
We correct for the degradation in our sensitivity due to anisoplanatism from the 2005 $H$-band data\footnote{Nominally, 
we adopt the scaling of Strehl ratio reduction vs. angular distance measured for Keck/NIRC2 from \citet{VanDam2006}.  
For comparison, we measured FWHM along the x and y directions of bright point sources at $r$ $>$ 10\arcsec{} 
in the 2005 data.  The performance reductions are in fair agreement with 
those noted by \citet{Kalas2008}.}.
Finally, we impute artificial point sources into registered images to measure and correct 
for flux losses from candidate companions due to PSF subtraction and PSF smearing from field rotation 
 \citep[e.g.][]{Lafreniere2007, Currie2010}.  
\section{Results}
\subsection{Reduced Images and Candidate Companions}
\begin{figure}
%\centering
\epsscale{1.1}
\plottwo{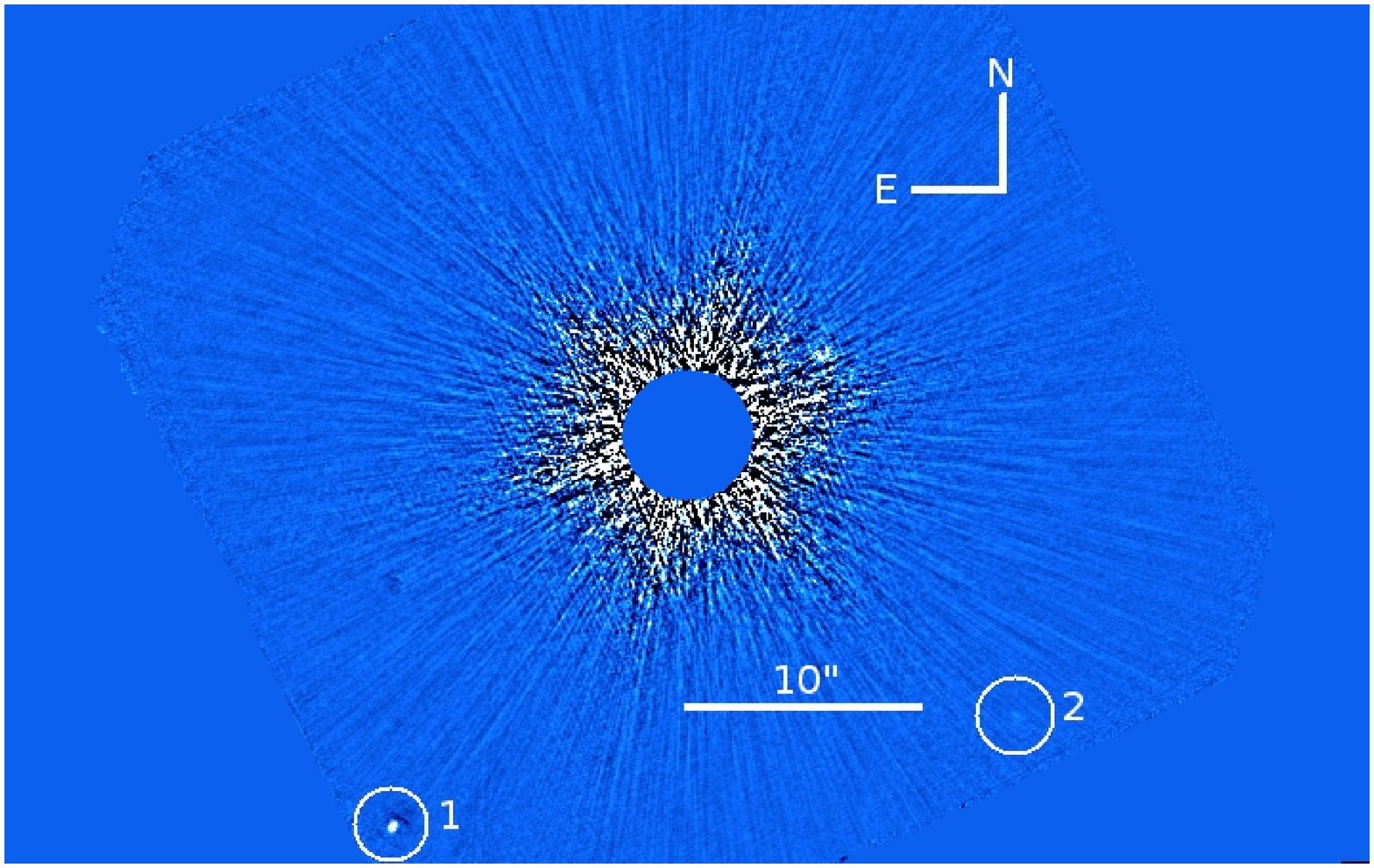}{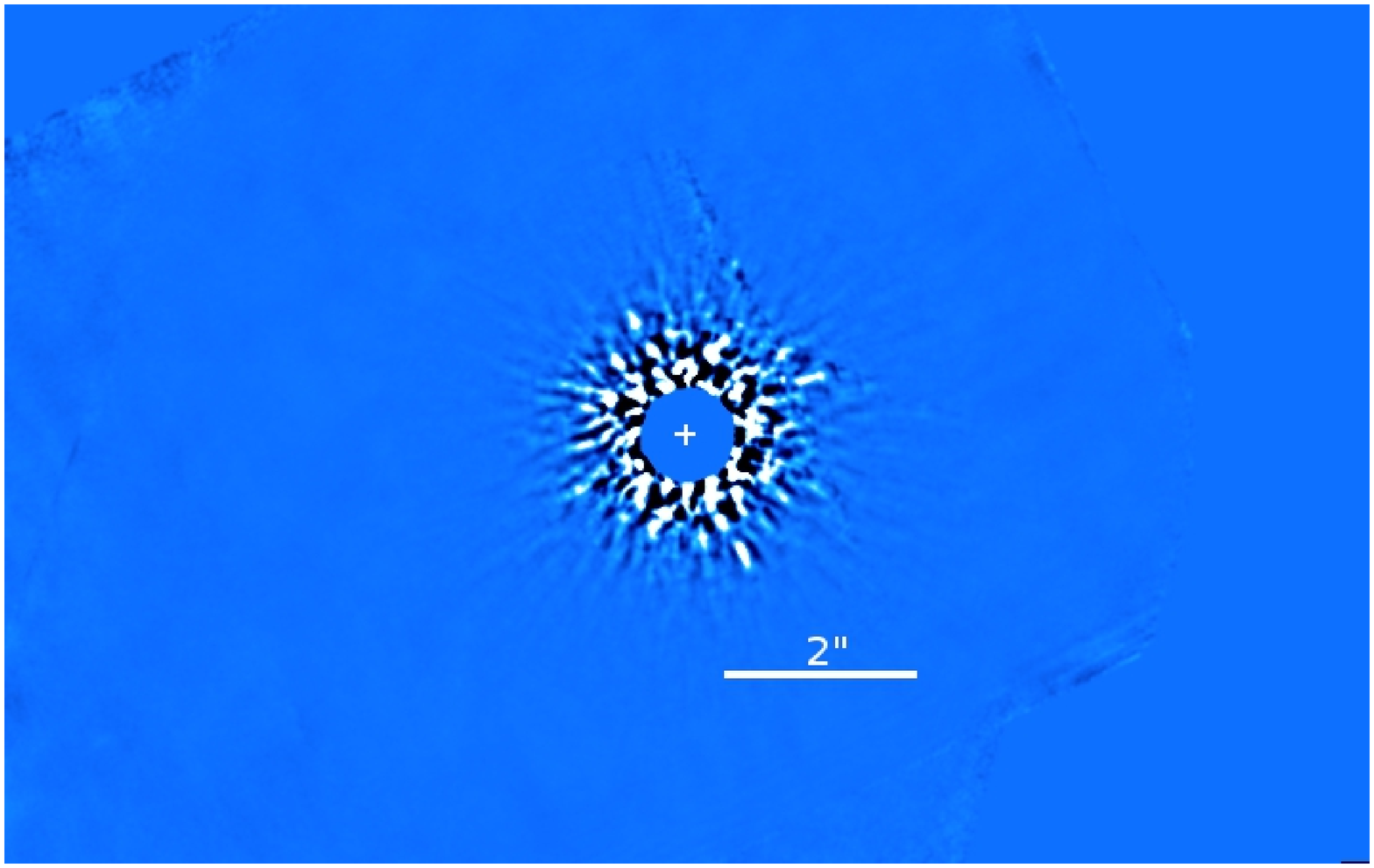}
\plottwo{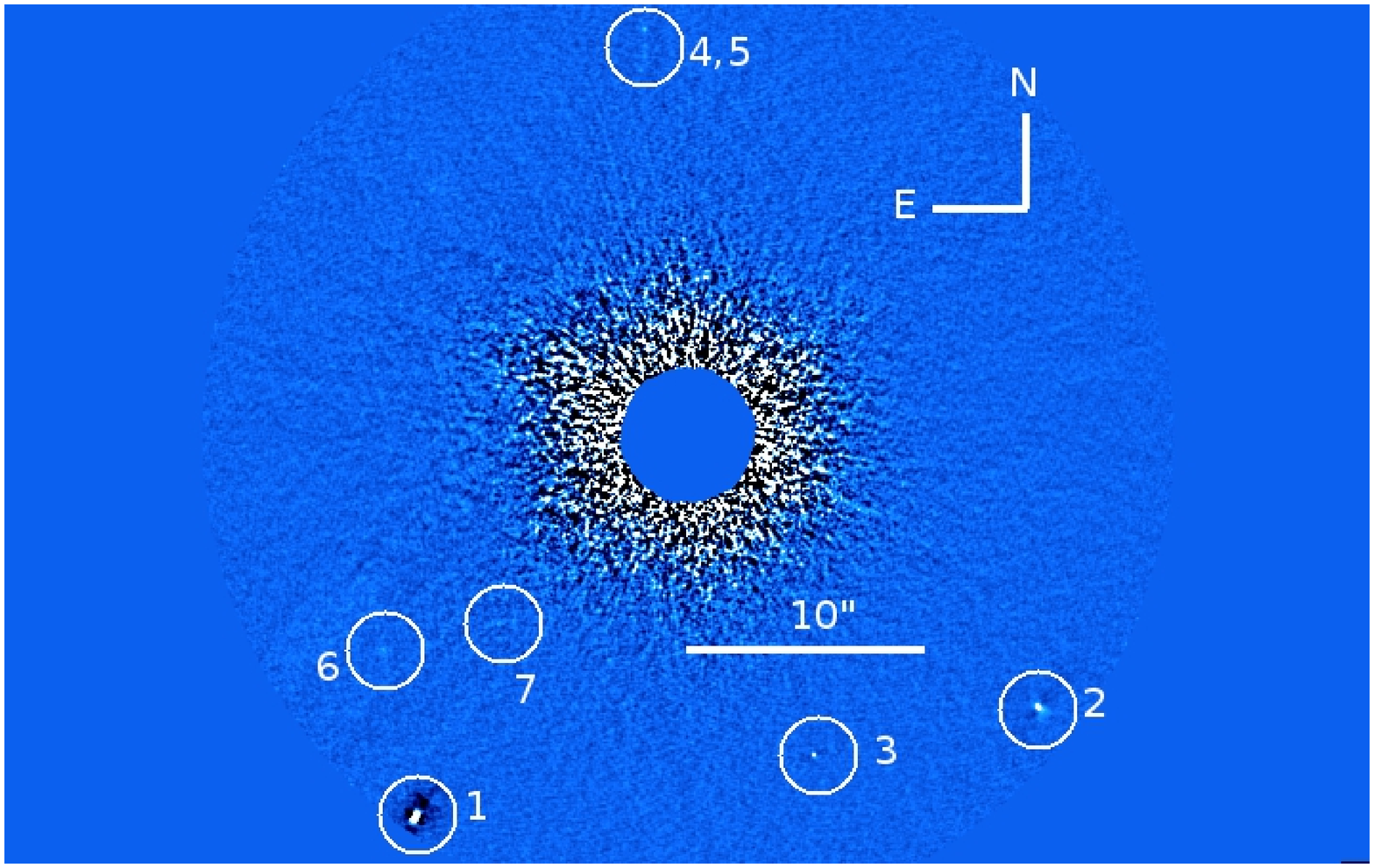}{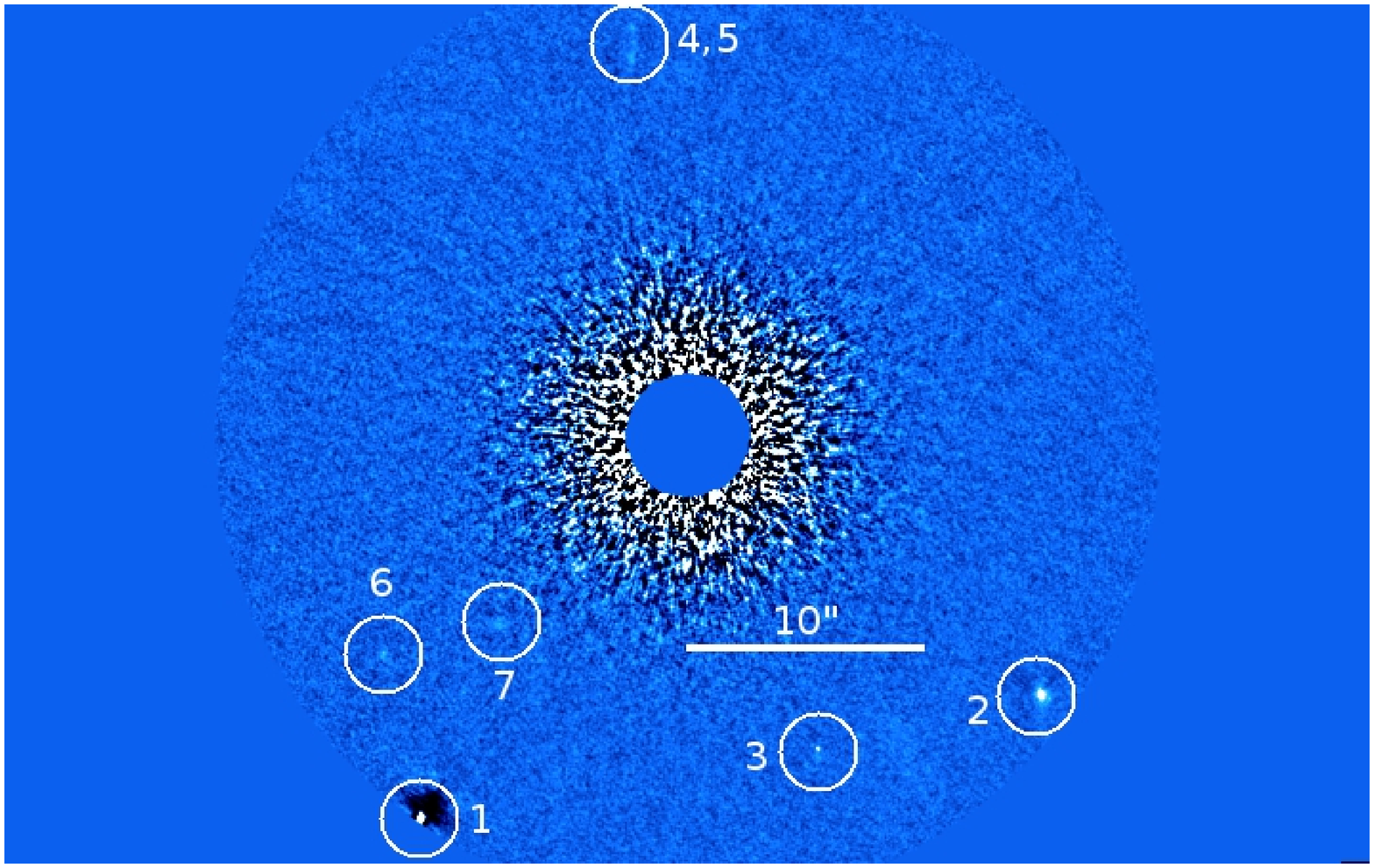}
\caption{Reduced Keck/NIRC2 images (counterclockwise from top-left): 2002 $H$ band data, July 2005 $H$ band 
data, October 2005 $H$ band data, and July--August 2003 $L^\prime$ data.  All images are rotated 
``north-up" and a horizontal bar denotes the images' spatial scales.  The color stretch in each image is 
defined to highlight the regions with significant speckle noise contamination.  We identify 
seven off-axis objects in the $H$ band data but none at $L^\prime$.}
\label{images}
\end{figure}

Figure \ref{images} shows the reduced images obtained for the 2002 $H$-band data (top-left), 
2003 $L^\prime$ data (top-right), the July/October 2005 $H$-band data (bottom panels). 
The inner working angles (IWA) for the 2002 and 
2005 $H$-band data are $\approx$ 2\farcs{}3--2\farcs{}5, or 
$\sim$ 17.5--20 $AU$ at Fomalhaut's distance.  
%Although the 2010 $H$-band 
%data achieve a significantly larger parallactic angle rotation, the IWA is much 
%larger ($r$ $\sim$ 4\farcs{}) since the observers took a single 30-second exposure 
%instead of 30 seconds with of shorter, coadded exposures and thus saturated 
%the star outside of the coronagraph.
The $L^\prime$ data are unsaturated outside of the coronagraph spot and 
thus are sensitive to companions with separations as small 
as 0\farcs{}5, or $\approx$ 4 AU.

We detect seven off-axis objects in total (Table \ref{candcomp}).
Though the 2002 $H$-band data are shallow and have only modest speckle 
suppression, we identify two wide-separation objects (labeled
as ``1" and ``2" in Figure \ref{images}, top-left panel).  Both 2005 $H$-band datasets 
(bottom panels) reveal 5 more objects, at least one, perhaps two of which 
 have small angular extents consistent with point sources given the PSF undersampling and anisoplanatic 
effects ($\theta$ $\lesssim$ 0\farcs{}1).  As Fomalhaut b might extended 
at red wavelengths \citep{Galicher2013}, the other sources we see 
could in principle be analogues.
%The October 2005 identify a 
%possible eigth object that appears spatially extended. 
At the age of Fomalhaut, the brightnesses of the candidate companions are consistent 
with $M$ $\sim$ 5--15 $M_{J}$ objects \citep{Baraffe2003,Spiegel2012}.
%The 2010 $H$-band data recover 
%each of these objects and reveal xxx additional ones.  
The $L^\prime$ data do not clearly reveal any candidate companions within 
5\farcs{} of the star (top-right panel).  We do not detect any 
point source at the location of the $\sim$ 4-$\sigma$ peak 
identified in \citet{Janson2012} ([E,N]\arcsec{} = 6.6, -8.7).

%We derive aperture photometry for each object detected on the $H$-band fields 
%within a 2-pixel radius, correcting for throughput losses and anisoplanatic 
%effects (Table \ref{candcomp}).  
%The $H$-band magnitudes range between $m(H)$ $\approx$ 16 and 22.  
%In principle the flux calibration assumes that the objects we 
%detect are point sources, not extended sources, whereas most objects 
%are very clearly extended.  

\begin{figure}
\plottwo{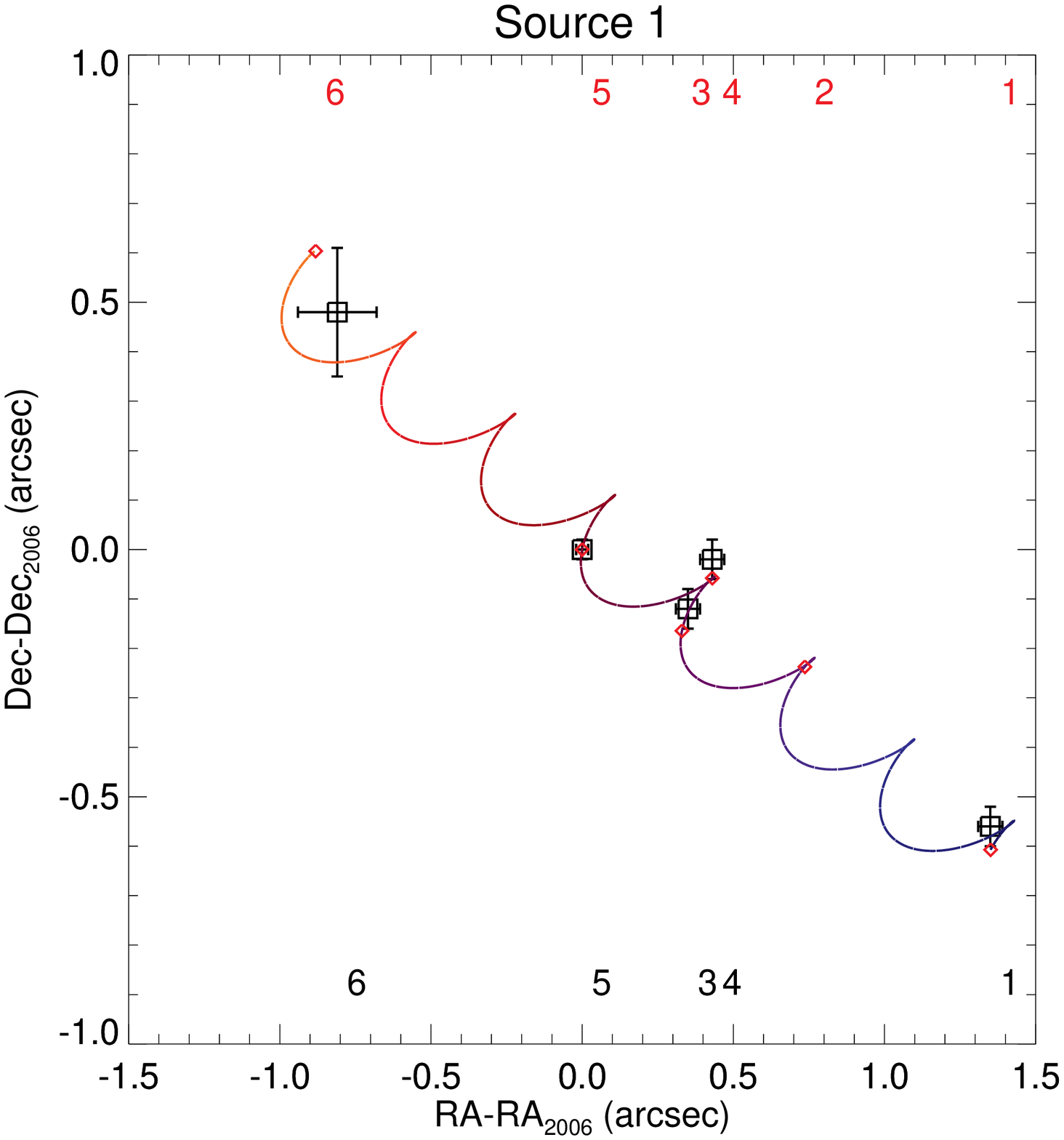}{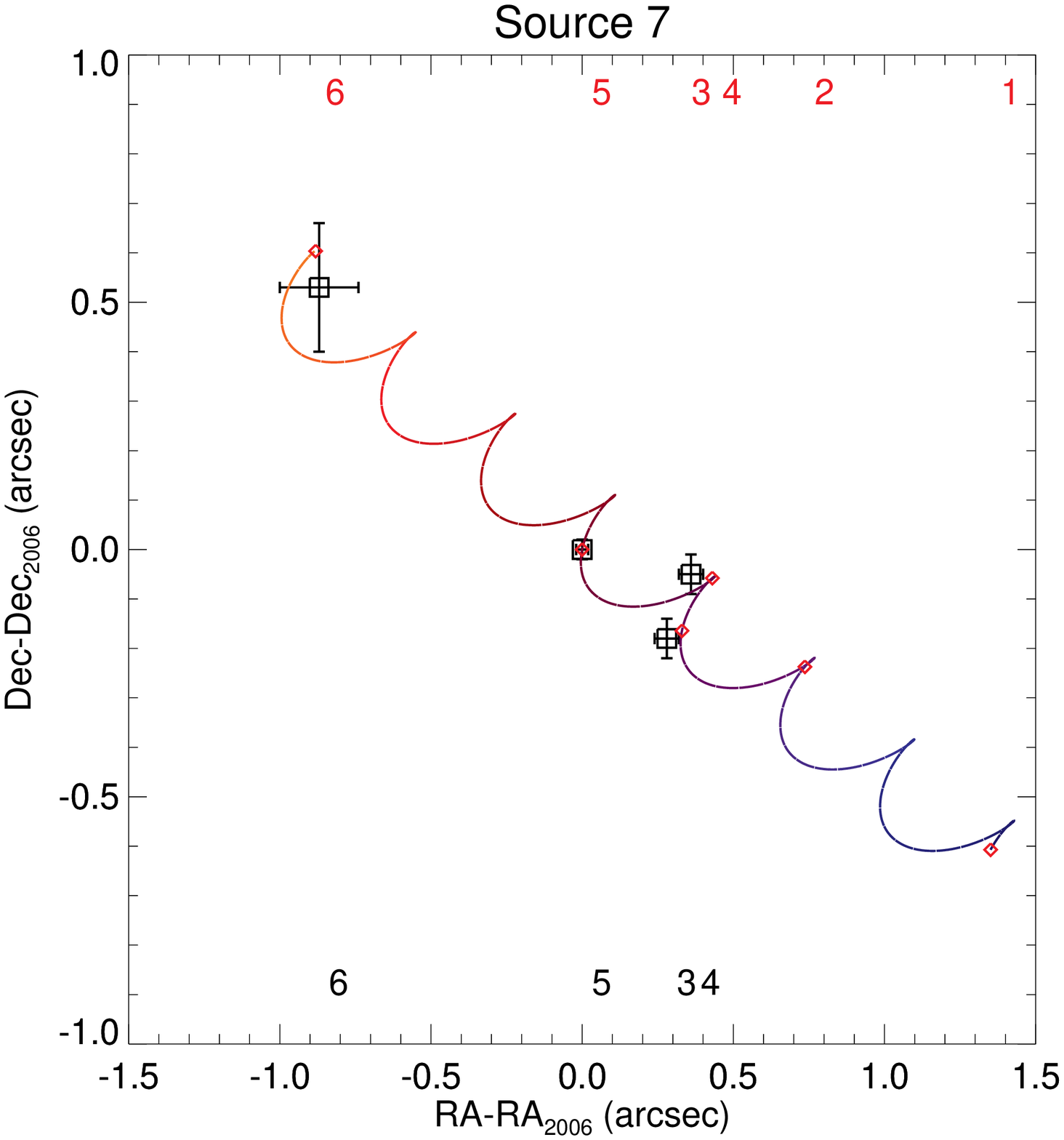}
\caption{Proper motion analysis for sources 1 and 7 listed in Table \ref{candcomp}: 
both plots compare positions to the 2006 $HST$ positions from \citet{Currie2012a}.  Open black 
squares with error bars denote our measurements; small red squares are the predicted 
positions for background stars.  Combining 
the Keck data with our reduction of the HST/ACS data \citep{Currie2012a} and WFC3 data (this work) 
shows that these and all other candidate companions (Table \ref{candcomp}) are 
background objects.}
\label{propmo}
\end{figure}

To establish whether these objects are bound companions, we compare their 2002 and 2005 
NIRC2 positions with those from 2004--2006 $ACS$ data reduced in \citet{Currie2012a} 
(Figure \ref{propmo}).  
We further added 2009 $HST/WFC3$ data with basic processing as described in \citet{Currie2012a} 
and $PSF$ subtraction as performed here for Keck/NIRC2 data.  
We adopt the narrow camera north position angle offset of -0.252$^{o}$, since there should be no difference here 
between the narrow and wide cameras (H. Fui, 2013, pvt. comm.): our analysis likewise did not identify clear evidence 
for an offset.  The seven objects have astrometry consistent 
with that of background objects: the extended sources are then likely 
background galaxies.  Sources 2, 3, and 5 are consistent with previously identified background objects 
in \citet{Currie2012a}; \citet{Galicher2013} also show that an object consistent with Source 3 is a background object and 
flag Source 5 in their images.  
\subsection{Limits on Additional Companions and the Near-IR brightness of Fomalhaut b}
\textbf{Contrast Limits} -- To derive upper limits on the brightness of Fomalhaut b and 
additional companions, 
we calculate the dispersion $\sigma$ in pixel values at a given angular separation \citep[c.f.][]{Marois2008b}
over an aperture area with a two (9.4) pixel diameter for $H$ ($L^\prime$) data 
to derive the formal 3-$\sigma$ and 5-$\sigma$ contrast and apparent magnitude limit.  
We combine the 2005 $H$-band data sets to enhance our point source sensitivity 
\footnote{A bound companion moves too little between epochs to affect our limits 
at $H$ band.  Given Fomalhaut b's astrometry \citep{Currie2012a,Galicher2013}, it should 
move $\approx$0\farcs{}02 between epochs, or $\sim$ half a NIRC2 pixel.   Given the 
PSF broadening at separations where $H$-band data are more sensitive than $L^\prime$, 
this effect is minimal.  Three of the four $L^\prime$ data sets were obtained within 2 days of one another.}.
%Limits we would derive
%assuming that the noise is gaussian, uncorrelated, dominated by background fluctuations, and scaling 
%as N = $\sigma$(r)$\sqrt{\pi r^{2}}$ \citep[c.f.][]{Kenworthy2013} 
%are roughly a factor of 1.8 and 3.8 times deeper at $H$ and $L^\prime$ than what we 
%report here, respectively\footnote{While residual speckle noise statistics post-processing 
%likewise converges to gaussian, the noise is correlated and the variance scales
%quadratically with $Flux_{speckle}$ \citep{Racine1999,Soummer2004}, such that $\sigma_{speckle}(r)$ 
%$\sim$ $I_{s(r)}\pi r_{aperture}^{2}$.}.  

\begin{figure}
\centering
\includegraphics[scale=0.45,trim=5mm 5mm 5mm 5mm,clip]{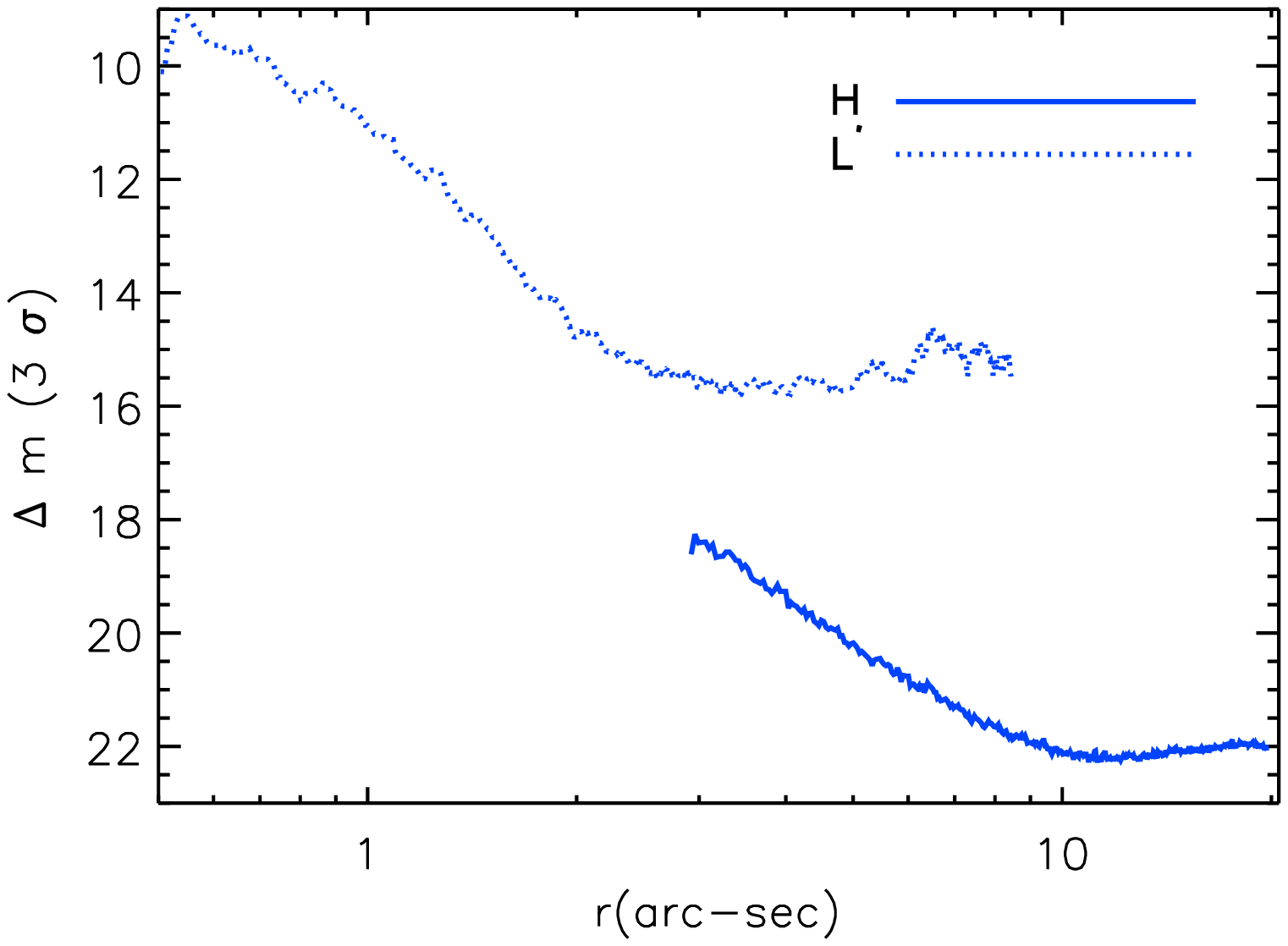}
\includegraphics[scale=0.4,trim=0mm 0mm 25mm 0mm,clip]{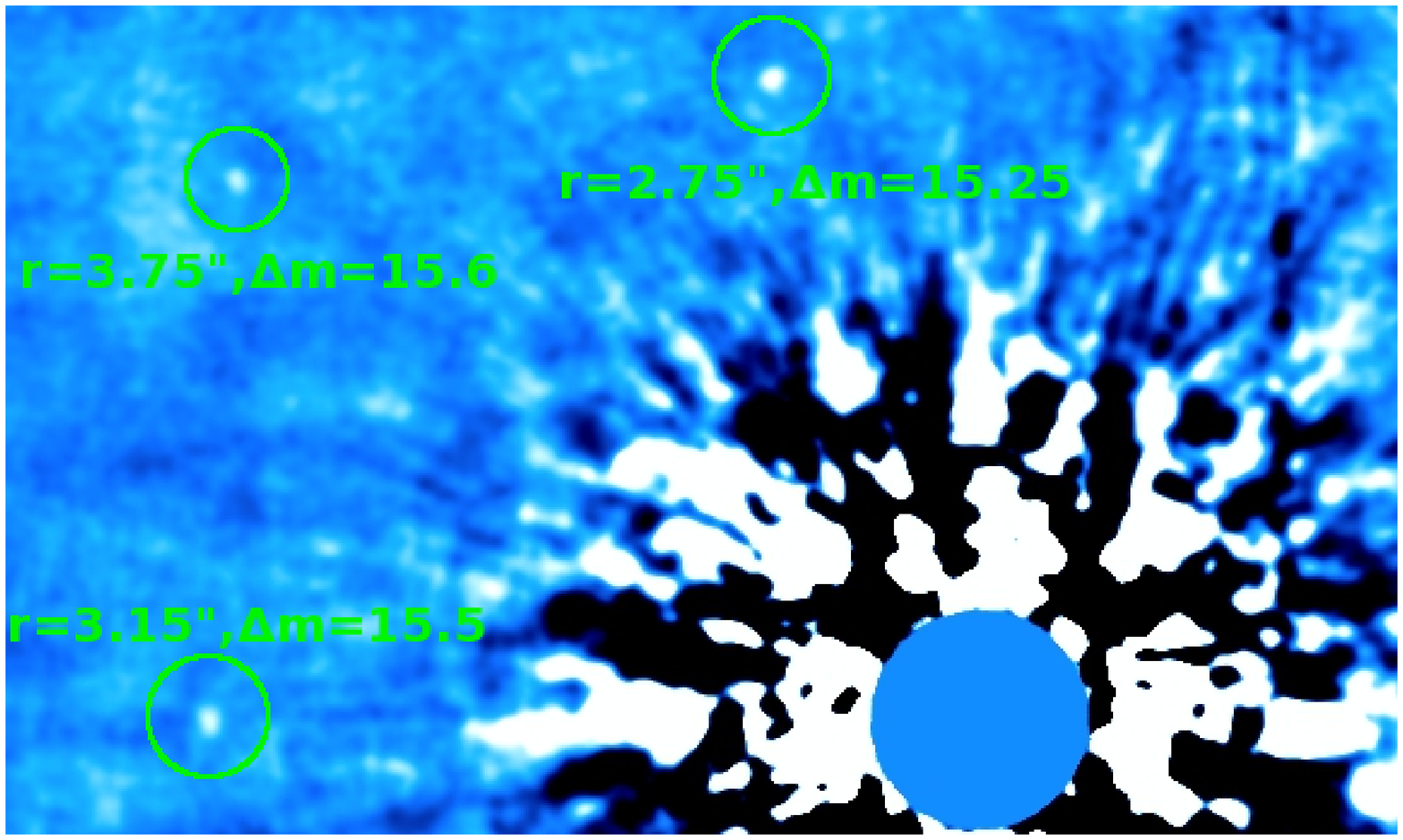}\\
\includegraphics[scale=0.45,trim=5mm 5mm 5mm 0mm,clip]{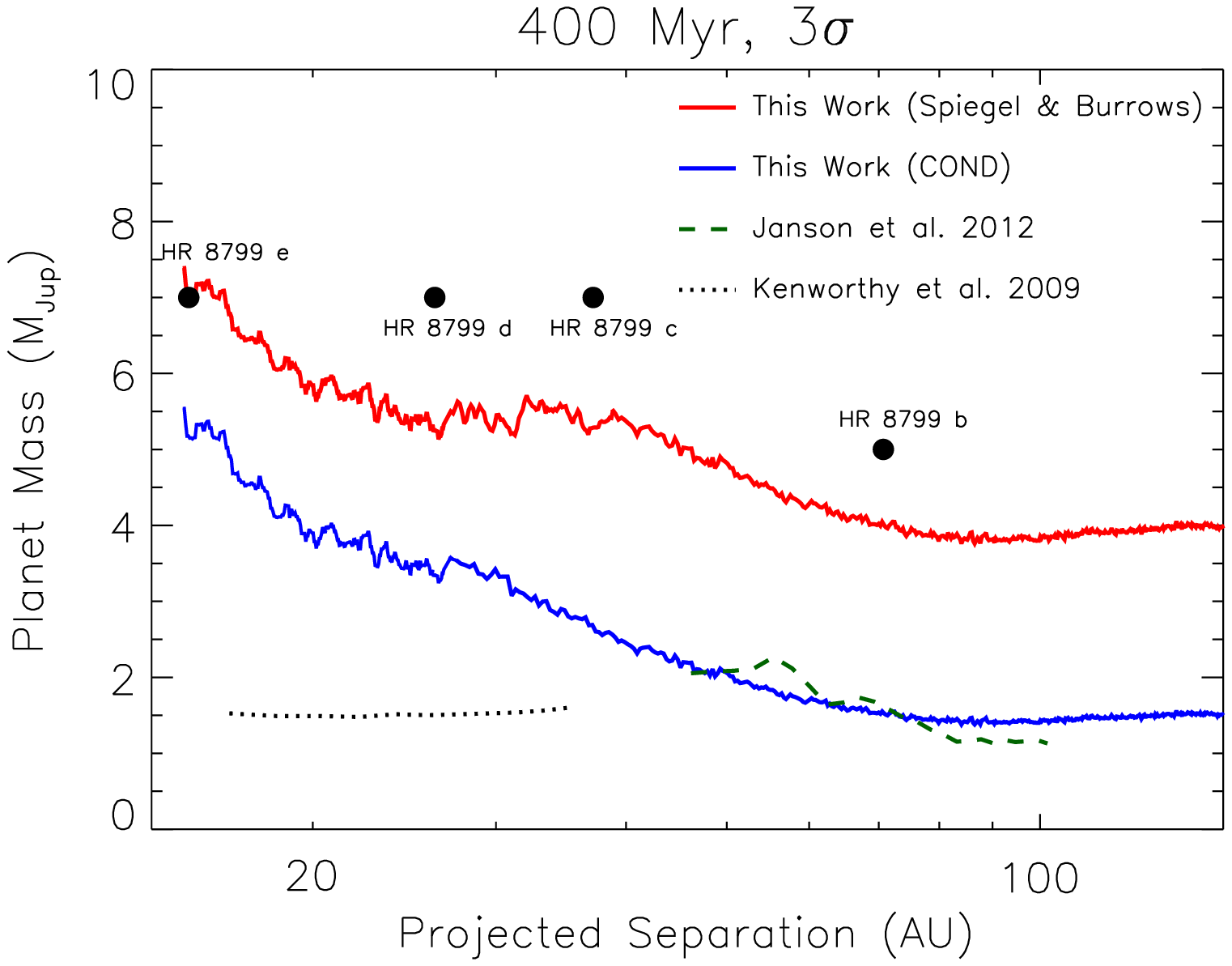}
\includegraphics[scale=0.45,trim=5mm 5mm 5mm 0mm,clip]{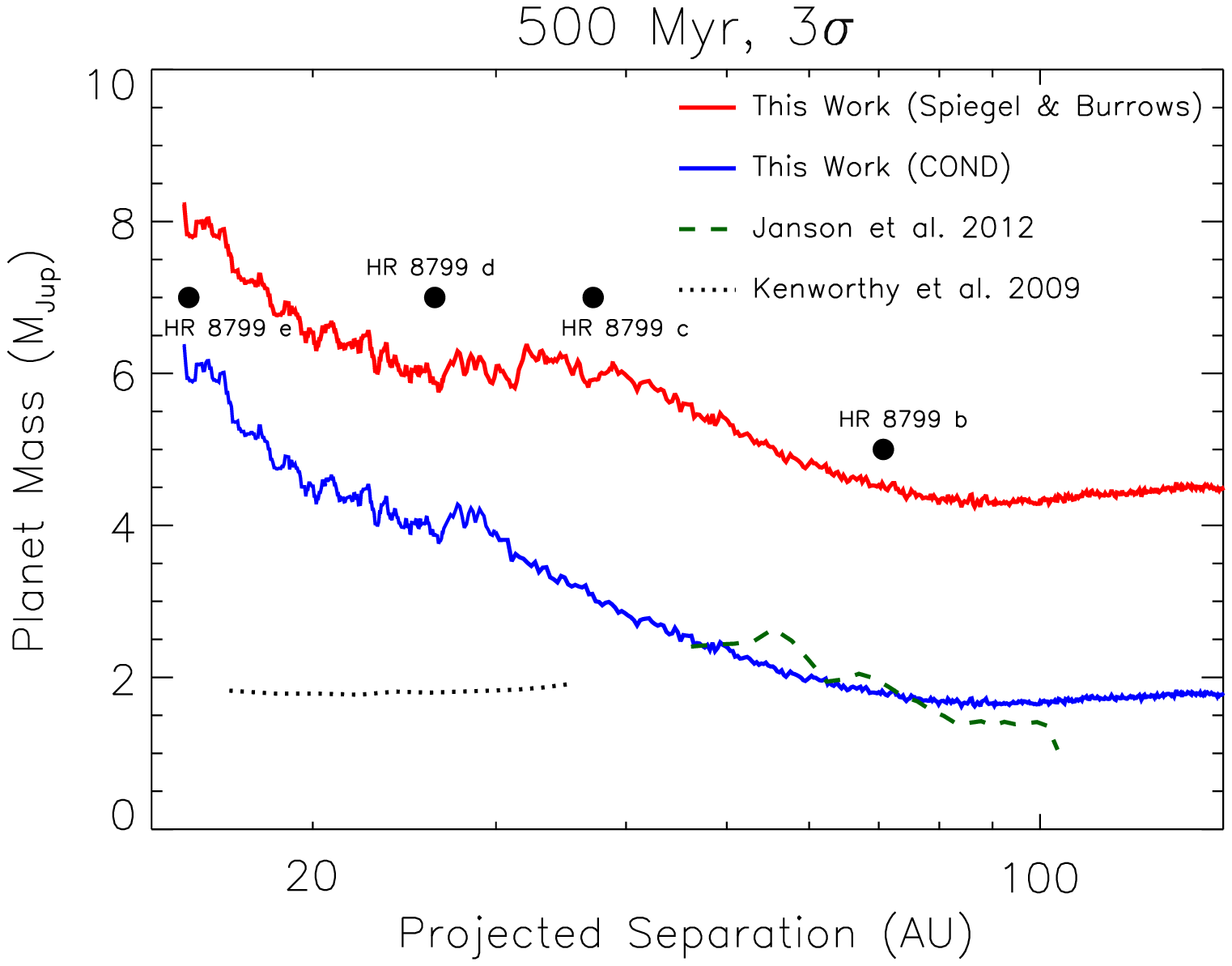}
\includegraphics[scale=0.45,trim=5mm 5mm 5mm 0mm,clip]{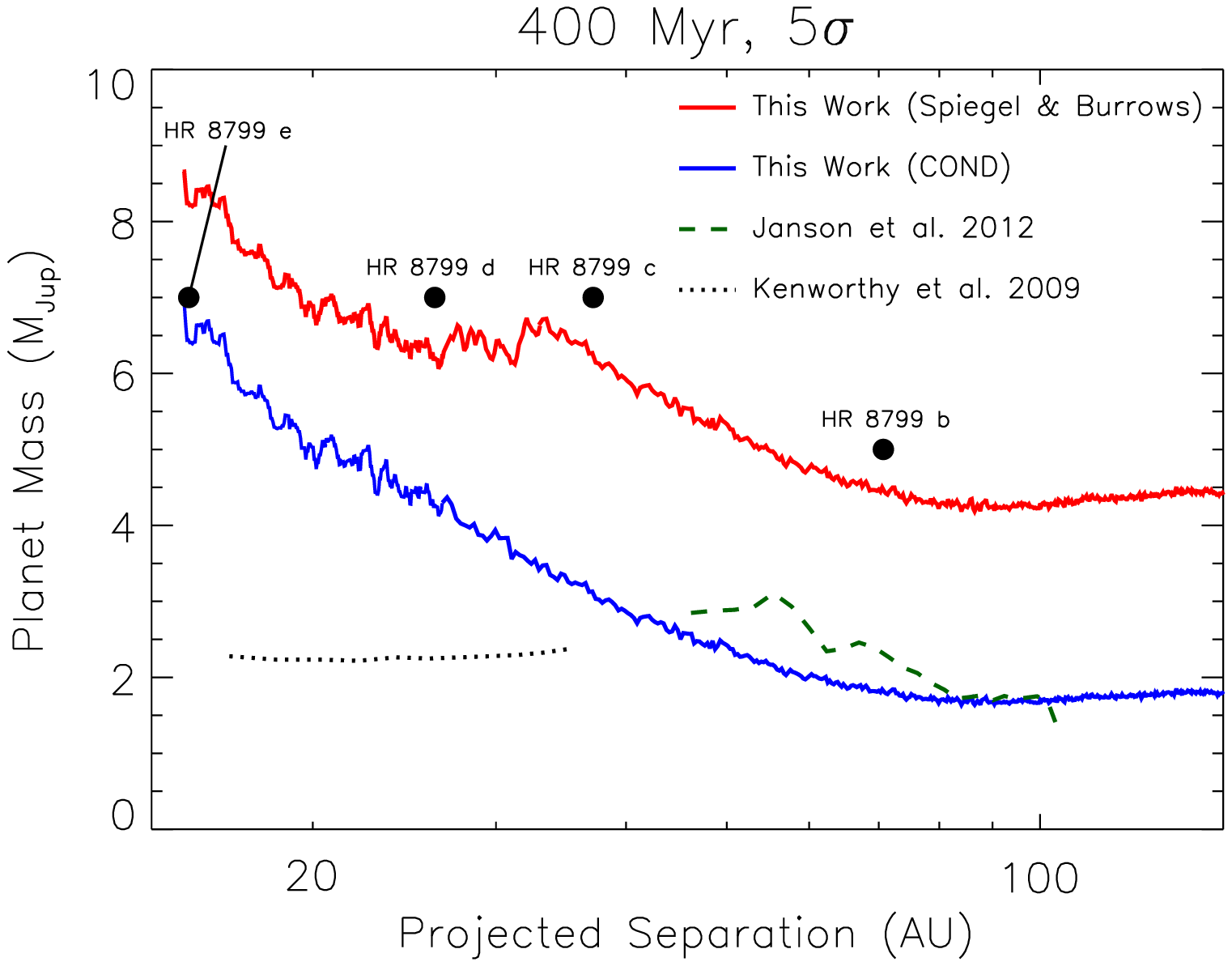}
\includegraphics[scale=0.45,trim=5mm 5mm 5mm 0mm,clip]{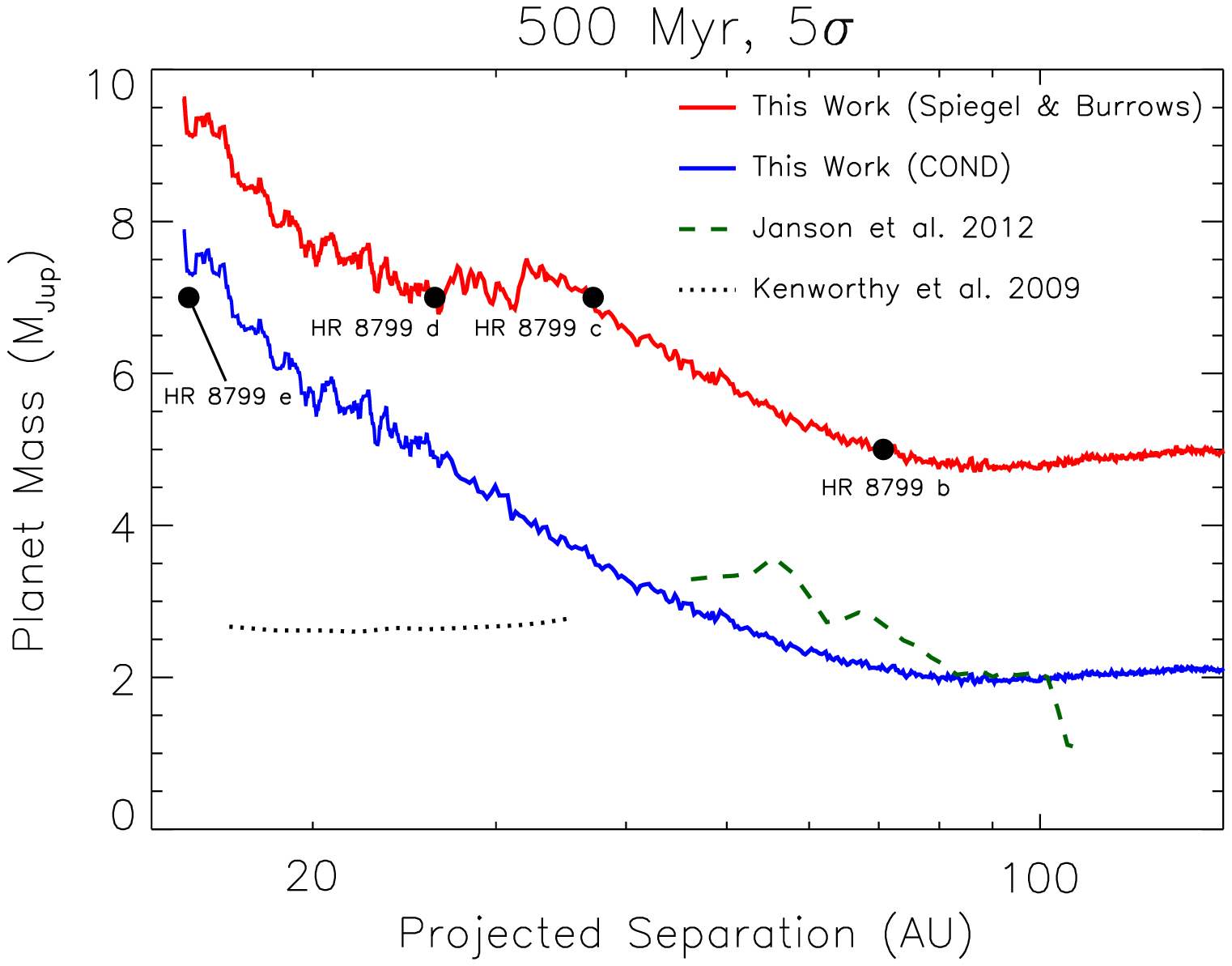}
\caption{(Top-left) Contrast limits (3-$\sigma$) and tests showing that 
we can detect planets at these limits (top-right).  (Middle and Bottom panels) 
Our 3$\sigma$ (middle panels) and 5$\sigma$ (bottom panels) mass limits 
assuming an age of 400 $Myr$ (left panels) and 500 $Myr$ (right panels).}
\label{conmasslim}
\end{figure}

Figure \ref{conmasslim} (top-left panel) displays our 3-$\sigma$ contrast limits at $H$ and $L^\prime$, where 
magnitude limits can be derived by adding 0.94 to each curve.  
The $L^\prime$ contrast limit is $\Delta$$L^\prime$ $\sim$ 9.3--11.2 
at the smallest separations ($r$ =0\farcs{}5--1\arcsec{}). 
%is deep enough to detect an object with HR 8799 c and d's contrast \citep[$\Delta$$L^\prime$ 
%$\sim$ 9.3--9.5][]{Marois2008,Currie2011a}.
%, while GJ 504 b ($\Delta$$H$ $\sim$ 16.4, $H$ = 20.1, 
%$r$ $\sim$ 2\farcs{}5, \citealt{Kuzuhara2013}) would likewise be detectable.
Contrast limits for $H$-band at $r$ $>$ 13\farcs{} and for $L^\prime$ at $r$ $>$ 5\farcs{} degrade 
 due to anisoplanatism and poorer sky coverage, respectively.  
The $L^\prime$ limits reach $\Delta$$L^\prime$ $\sim$ 15.7 at $r$ $\sim$ 3\arcsec{}--5\arcsec{}.  
The $H$-band limits reach $\Delta$$H$ $\sim$ 22.2 or m($H$) $\approx$ 23.15 at 
a range of separations enclosing that of Fomalhaut b and the possible point source reported 
in \citet{Janson2012} (10\farcs{}75-13\arcsec{}).  
We verified that we can detect planets at our contrast limits.  
For example, in Figure \ref{conmasslim} (top-right panel) 
 we input and recover three artificial point sources into the $L^\prime$ data with brightnesses equal to 
our estimated 3-$\sigma$ contrast limits at 2\farcs{}75, 3\farcs{}15, and 3\farcs{}75 
($\Delta$$L^\prime$ $\sim$ 15.25, 15.5, and 15.6).  

\textbf{Planet Detection Limits} -- To derive mass upper limits from our magnitude limits, 
we adopt the Hipparcos-measured 
distance of 7.7 $pc$, an age of $t$ = 400--500 $Myr$ \citep{Mamajek2012},
and the \citet{Spiegel2012} (``hybrid clouds", solar metallicity) 
and the \citet{Baraffe2003} COND hot-start planet evolution models
to map between $H$ and $L^\prime$ magnitudes and planet mass (Figure \ref{conmasslim}).  
At Fomalhaut's age, planet luminosities are independent of 
the initial entropy \citep[c.f.][]{Spiegel2012}.
%While previous studies have compared planet mass limits by varying the metallicity and 
%cloud presence/absence within a given atmosphere model \citep{Janson2012,Currie2012a}, 
%The \citeauthor{Spiegel2012} and COND models give very different predictions for 
%planet brightnesses at 1.25--4 $\mu m$ \citep[c.f.][]{Kenworthy2013}.
%  Thus, we derive separate mass limits for each model.
We plot a single curve for each planet cooling model, combining together the $H$ and $L^\prime$ limits.
The $L^\prime$ data are deeper at $r$ $\lesssim$ 5\farcs{}2 (4\arcsec{}) 
for the \citeauthor{Spiegel2012} (COND) models.
We focus on 2\arcsec{}--20\arcsec{} ($r_{proj}$ $\sim$ 15--150 $AU)$ where we have 
continuous coverage; \citet{Kenworthy2013} and \citet{Meshkat2013} explore planet mass limits at smaller separations.
% from 
%data more sensitive at those locations.

The middle and bottom panels of Figure \ref{conmasslim} display our mass sensitivity 
limits at the 3-$\sigma$ (middle) and 5-$\sigma$ (bottom) level assuming 
ages of 400 $Myr$ (left) and 500 $Myr$ (right) assuming the COND (blue lines) 
and Spiegel \& Burrows (red lines) models.  For the COND models, our 
3-$\sigma$ upper limits rule out planets with masses $\gtrsim$ 5--6 $M_{J}$ at 
a projected separation of 15 AU, $\sim$ 2.5 $M_{J}$ at 40 AU and 1.5--2 $M_{J}$ at $r$ $>$ 70 AU. 
For the \citet{Spiegel2012} models, the corresponding limits are 7--8 $M_{J}$ at 15 AU, 
5--6 $M_{J}$ at 40 AU, and $\sim$ 4.5 $M_{J}$ at $r$ $>$ 70 AU.  
At the 5-$\sigma$ level, our limits are 7--8 $M_{J}$ (8--9 $M_{J}$) at 
15 AU, $\sim$ 3 $M_{J}$ (5.5--6 $M_{J}$) at 40 AU and 1.75--2 $M_{J}$ (4.5--5 $M_{J}$) at 
$>$ 70 AU assuming the COND (Spiegel \& Burrows) models.
If the possible point source 
identified in \citet{Janson2012} is real, then our 5-$\sigma$ limits suggest it must have a
mass less than $\sim$ 1.75--2 $M_{J}$ (4.5--5 $M_{J}$) assuming the COND (Spiegel \& Burrows) models.

%\textbf{Planet Detection Limits Combined with Previous Work}--
To determine the range of planet masses at $r_{proj}$ = 15--150 $AU$ consistent with non-detections reported in the literature, 
we combine our results with those from \citet{Janson2012} and \citet{Kenworthy2009}\footnote{
For simplicity, we do not include the \citet{Kenworthy2013} limits as they define $SNR$ (and thus contrast) differently 
than here and in \citet{Janson2012}.  \citet{Kenworthy2009} state that their data are background limited, 
not speckle noise limited, over the separations we focus on here ($r$ $>$ 2\farcs{}0).}.
Planets orbiting Fomalhaut comparable in mass and projected separation to HR 8799 bcde ($r_{proj}$ $\sim$ 15--70 $AU$) 
 should have been detected.
Over most (all) position angles, HR 8799 bcd-like (HR 8799 bc-like) planets in a ring-nested orbit 
\citep[$i$ $\sim$ 66$^{o}$,][]{Kalas2005,Currie2012a} are detectable in at least one of the data sets.
%%REVISE!!!
At all projected separations wider than $\sim$ 45 $AU$, we rule out a $M$ $\ge$ 3 $M_{J}$ planet, 
comparable to the mass of an object modeled as sculpting Fomalhaut's debris ring 
\citep{Chiang2009}.
%and 
%the IR upper limits derived for Fomalhaut b \citep{Chiang2009,Currie2012a,Janson2012,Galicher2013}.

\textbf{New IR Detection Limits for Fomalhaut b} -- 
We derive a new 3-$\sigma$ upper limit for Fomalhaut b of m($H$) $\sim$ 23.15, slightly deeper than that derived 
by \citet{Kalas2008} using the same data, which excludes planets with masses greater than 
4--4.5 $M_{J}$ from the \citet{Spiegel2012} models and 1.5--2 $M_{J}$ from the COND models 
for an age of 400--500 $Myr$.  
These limits are comparable to those derived from Spitzer/IRAC, Subaru/IRCS, and 
$HST$/$WFC3$ \citep{Janson2012,Currie2012a,Galicher2013} and are typically a few tenths of 
Jupiter-masses lower than reported by \citet{Kalas2008} using a different metric for determining 
contrast.  No current data are sensitive 
enough to rule out a Jupiter mass for Fomalhaut b, let alone a Saturn.

\section{Discussion}
%We present new, deep limits on thermal emission from planetary companions orbiting Fomalhaut, using 
%seven archival Keck/NIRC2 data sets taken between 2002 and 2005.  While we identify seven
%faint off-axis objects on the field at $H$ band, all are consistent with being background objects, 
%not bound companions.  
%Combining planet detection limits with those from other recent studies, 
%Depending on which model we use to map between planet brightness, mass, and age,
%our data rule out planets/brown dwarfs with masses greater than 12--14 $M_{J}$ 
%at 8--15 AU, planets with masses greater than 3.5--6 $M_{J}$ at $r$ $>$ 15 AU and 
%just exterior to Fomalhaut's inner dust belt, and 1.5--4 $M_{J}$ at 50--100 AU. 
%Fomalhaut b remains undetected in the infrared, 
%even using advanced PSF subtraction techniques\footnote{We also reduced
 %deeper $H$-band NIRC2 Fomalhaut public data sets obtained between 2009 and 2011. 
%  The results presented in this paper are not undermined, although 
%we obtain slightly better upper limits on Fomalhaut b's brightness.  Because this paper
%is not primarily focused on new Fomalhaut b upper limits, we defer the reporting of 
%such limits to work lead by other authors (M. Fitzgerald 2013, in prep.)}.

Given our mass detection limits, Fomalhaut's planetary system is now clearly identified 
as being different than at least some other planetary systems identified from direct imaging.  
Fomalhaut has apparently failed to form and retain planets at $r_{proj}$ $>$ 15 AU comparable 
in mass to directly-imaged planets around HR 8799.  
%HR 8799 has a very luminous debris disk, as do $\beta$ Pic 
%and HD 95086, which have confirmed and candidate planets \citep{Lagrange2010,Rameau2013}.
%The $\beta$ Pic disk even shows signatures of planet dynamical sculpting; HR 8799's show 
%evidence for multiple belts \citep{Golimowski2006,Su2009}.
%Fomalhaut shares these debris disk characteristics \citep{Kalas2005,Su2013}, yet does not have 
%planets with comparable masses ($M$ = 5--15 $M_{J}$, $a$ $\sim$ 15--70 $AU$).  
However, the mass limits we derive near the disk gap edge at most azimuthal angles are 
significantly deeper than those derived for debris disk-bearing stars in many direct imaging 
surveys \citep[e.g.][]{Janson2013}.  Although Fomalhaut's planet inventory is different than 
that of HR 8799, it may be no different than the typical star with a luminous debris disk.  
Conversely, a Fomalhaut b-like object at typical distances of these disks ($d$ $\approx$ 30-50 $pc$) 
would probably be undetectable with any currently-operating instrument.
While a planet with a mass/separation comparable to the candidate around HD 95086 
 \citep{Rameau2013} would have been detectable, 
we cannot rule out a $\beta$ Pic b analogue from our data \citep{Lagrange2010}.

Even though Fomalhaut's debris ring has been modeled as being sculpted by a super-Jovian mass 
planet, it is still unclear what determines the ring's structure.  
%Although \citet{Chiang2009} 
%considered a 3 $M_{J}$ planet to be a plausible explanation for the ring structure, 
 Planets with masses down to 0.5 $M_{J}$ can sculpt the ring in the \citet{Chiang2009} simulations; 
 perhaps even lower-mass planets located closer to the ring could likewise sculpt it.
Planets with masses of 0.5--3 $M_{J}$ and cold atmospheres at Fomalhaut's age are likely as faint as 
$M_{H}$ $\sim$ 22--30 \citep[][]{Baraffe2003,Spiegel2012} and thus will be undetectable 
within the control radius ($r$ $\lesssim$ 1\farcs{}0) of next-generation instruments like 
\textit{SCExAO}, \textit{GPI}, and \textit{SPHERE} \citep{Martinache2009,Macintosh2008,
Beuzit2008}, although a thermal infrared system like $NaCo$ with the \textit{Annular Groove Phase Mask} \citep{Mawet2013}
or $LBTI/LMIRCam$ \citep{Wilson2008}
might image slightly more massive companions.  Terrestrial mass ``shepherding planets" \citep{Boley2012}, 
which would likewise be undetectable, could also truncate the ring. 
The ring structure could even have non-planet origins \citep{Lyra2013} if the 
ring's gas-to-dust ratio is near unity.

%Fomalhaut's advanced age means that any hitherto undetected planet is likely to be 
%very cold and thus far brighter in the mid infrared than near infrared \citep[c.f.][]{Kuzuhara2013}.  
%While the upcoming generation of ground-based extreme adaptive optics imagers may be unable 
To shed further substantial light on Fomalhaut's planet inventory in the next 5--10 years, 
high-contrast mid-IR imaging with NIRCAM on
the \textit{James Webb Space Telescope} will be key \citep[][]{Krist2007}.  
NIRCAM should achieve 
a 5-$\sigma$ contrast of $>$ 10$^{8}$ at $r$ $>$ 5\arcsec{}, making even Saturn-mass planets 
detectable at $r$ $\gtrsim$ 35 AU \citep[c.f.][]{Baraffe2003,Spiegel2012}. 
A NIRCAM non-detection of thermal emission from Fomalhaut b may limit its mass to that of a (Super)-Neptune, a 
range more consistent with models of circumplanetary dust production from collision 
between satellites \citep{Kennedy2011}.  Fomalhaut b could be the 
first of many exoplanets studied with \textit{JWST}.  NIRCAM's extreme mid-IR sensitivity 
will make older ($t$ $\gtrsim$ 250--500 $Myr$) and/or colder ($T_{eff}$ $\lesssim$ 400--500 $K$) 
Jupiter-mass planets at moderate/wide ($r$ $\sim$ 1--10\arcsec{}) separations 
around the nearest stars imageable for the first time \citep[][]{Beichman2010}, 
complementing $GPI$/$SPHERE$/$SCExAO$ and later 20-30 m class telescope 
instrumentation \citep[e.g.][]{Hinz2012} focused on imaging jovian planets around 
younger, more distant stars and/or planets at smaller angular separations.

\acknowledgements 
We thank Timothy Rodigas and Mickael Bonnefoy for draft comments; 
Hai Fu and Stanimir Metchev for discussions regarding the NIRC2 astrometric 
calibration; Christian Marois for discussions regarding observing techniques; 
%Edwin Kite, Wladimir Lyra, and Marc Kuchner for discussions regarding planet/non-planet debris ring 
%sculpting; 
Markus Janson and Matthew Kenworthy for providing their planet detection limits; 
Chas Beichman for helpful notes on JWST/NIRCAM performance;
and Ray Jayawardhana for other helpful comments.
This research has made use of the Keck Observatory Archive (KOA), which is operated by the 
W. M. Keck Observatory and the NASA Exoplanet Science Institute (NExScI), under contract 
with the National Aeronautics and Space Administration. 

{}

\begin{deluxetable}{lcllccccccc}
\tablecolumns{8}
\tablecaption{Observing Log}
\tiny
\tablehead{{UT Date}&{Mode}&{Camera}&{$R_{mask}$}&{$IWA$}&{Filter}&{$t_{int}$}&{$N_{images}$}&{$\Delta$PA} &\\
{} & {}&{} &{ (\arcsec{})} &{ (\arcsec{})} &{}&{(s)}&{}&{(degrees)}}
\startdata
2002-08-21 & Classical & Wide& 1.0 & 2.3 & $H$ & 60 & 21 & 14.8\\
2003-07-27 & Classical & Narrow&0.5 & 0.5 & $L^\prime$ & 45 & 50 & 35.3\\
2003-07-28 & Classical & Narrow&0.5 & 0.5 & $L^\prime$& 35 & 50 & 28.1\\
2003-07-29 & Classical & Narrow&0.5 & 0.5 & $L^\prime$& 46 & 50 & 36.1\\
2003-08-19 & Classical & Narrow&0.5 & 0.5 & $L^\prime$ & 40 & 50 & 34.1\\
2005-07-17 & ADI & Wide&1.0 & 2.5&$H$ & 30 & 110 & 39.2\\
2005-10-21 & ADI & Wide&1.0 & 2.5&$H$ & 30 & 194 & 50.8\\
%2009-07-19 & ADI & Wide&-  & 0.5 & $L^\prime$& 5 & 426 & 44.4\\
%2010-07-10 & ADI & Wide&1.0 & 5.0 & $H$ & 30 & 405 & 71.5\\
 \enddata
%\tablecomments{}
\label{obslog}
\end{deluxetable}

\begin{deluxetable}{lcllccccccc}
\tiny
\tablecolumns{8}
\tablecaption{Candidate Companions Identified from Ground-Based Imaging}
\tiny
\tablehead{{Number}&{Epoch}&{m$_{H}$}&{[E,N] (\arcsec{})}&{Extended?}&{Status?}}
\startdata
1 & 2002-08-21 & & [12.36,-16.26] $\pm$ 0.04\\
  & 2005-07-17 & $<$ 16.9 $\pm$ 0.2 &[11.36,-15.82] $\pm$ 0.04 & yes&bckg\\
  %& 2005-07-17 & $<$ 16.3 $\pm$ 0.2 &[11.36,-15.82] $\pm$ 0.04 & yes&bckg\\
  & 2005-10-21 &  & [11.44,-15.72] $\pm$ 0.04\\
  & 2006-07-16 & &[11.01,-15.70] $\pm$ 0.03\\
  & 2009-11-16 & &[10.20,-15.22] $\pm$ 0.13\\
\\
%  & 2009-07-10 &\\
2 & 2002-08-21 & & [-13.64,-11.73] $\pm$ 0.04\\
  & 2005-07-17 & $<$ 20.8 $\pm$ 0.2 &[-14.57,-11.34] $\pm$ 0.04 & yes&bckg\\
  %& 2005-07-17 & $<$ 20.4 $\pm$ 0.2 &[-14.57,-11.34] $\pm$ 0.04 & yes&bckg\\
  & 2005-10-21 & & [-14.51,-11.20] $\pm$ 0.04\\
  & 2006-07-16 &&[-14.96,-11.05] $\pm$ 0.03\\
  & 2009-11-16 & &[-15.78,-10.55] $\pm$ 0.13\\
\\
%  & 2009-07-10 &\\
3 & 2004-09-26& &[-4.86,-13.29] $\pm$ 0.03\\
  & 2005-07-17 & 21.5 $\pm$ 0.2 &[-5.25,-13.31] $\pm$ 0.04 & no&bckg\\
  %& 2005-07-17 & 20.9 $\pm$ 0.2 &[-5.25,-13.31] $\pm$ 0.04 & no&bckg\\
  & 2005-10-21 & & [-5.16,-13.21] $\pm$ 0.04\\
  & 2006-07-16 & &[-5.59,-13.11] $\pm$ 0.03\\
  & 2009-11-16 & &[-6.41,-12.59] $\pm$ 0.13\\
\\
%  & 2009-07-10 &\\
4 & 2005-07-17 & $<$ 22.2 $\pm$ 0.2 &[1.78,16.93] $\pm$ 0.04 & yes?&bckg\\
%4 & 2005-07-17 & 21.8 $\pm$ 0.2 &[1.70,16.19] $\pm$ 0.04 & yes?&bckg\\
  & 2005-10-21 & & [1.88,17.10] $\pm$ 0.04\\
  & 2009-11-16 & & [0.67,17.60] $\pm$ 0.13\\
\\
%  & 2009-07-10 &\\
5 & 2005-07-17 & $<$ 22.7 $\pm$ 0.3 &[1.77,15.79] $\pm$ 0.08 & yes&bckg\\
  & 2005-10-21 & & [1.82,16.01] $\pm$ 0.08\\
  & 2009-11-16 & & [0.67,16.56] $\pm$ 0.13\\
\\
%  & 2009-07-10 &\\
6 & 2005-07-17 & $<$ 22.2 $\pm$ 0.3 & [12.72,-8.83] $\pm$ 0.04& yes&bckg\\
%6 & 2005-07-17 & $<$ 22.3 $\pm$ 0.3 & [12.72,-8.83] $\pm$ 0.04& yes&bckg\\
  & 2005-10-21 & & [12.80,-8.81] $\pm$ 0.04\\
  & 2009-11-16 & & [11.58,-8.12] $\pm$ 0.13\\
\\
%  & 2009-07-10 &\\
7 & 2005-07-17 & $<$ 22.4 $\pm$ 0.3 & [7.90,-7.83] $\pm$ 0.04& yes&bckg\\
  & 2005-10-21 & & [7.98,-7.70] $\pm$ 0.04\\
  & 2006-07-16 & & [7.62,-7.65] $\pm$ 0.06\\
  & 2009-11-16 & & [6.75,-7.12] $\pm$ 0.13\\

 \enddata
%\tablecomments{To derive photometry and determine whether any object 
%is extended, we focus only on the best-quality data (the July 2005 Keck data).}
\tablecomments{
%We perform photometry within a two pixel aperture radius to reduce 
%our measurement uncertainties
For extended objects, the ``$<$" denotes that the magnitude is an upper limit.  
}

\label{candcomp}
\end{deluxetable}

%\begin{deluxetable}{lcllccccccc}
%\setlength{\tabcolsep}{0pt}
%\tablecolumns{8}
%\tablecaption{Candidate Companions Identified from Ground-Based Imaging}
%\tiny
%\tablehead{{Number}&{Epoch}&{m$_{H}$}&{[E,N] (\arcsec{})}&{Extended?}&{Status?}}
%\startdata
%1 & 2002-08-21 &\\
%  & 2005-07-17 & $<$ 16.3 $\pm$ 0.2 &[-11.02,-16.18] & yes&bckg\\
  %& 2005-10-21 & \\
%  & 2009-07-10 &\\
%2 & 2002-08-21 &\\
  %& 2005-07-17 & $<$ 20.4 $\pm$ 0.2 &[14.86,-11.04] & yes&bckg\\
 % & 2005-10-21 &\\
%  & 2009-07-10 &\\
%3 & 2005-07-17 & 20.9 $\pm$ 0.2 &[5.55,-13.23] & no&bckg\\
%  & 2005-10-21 &\\
%  & 2009-07-10 &\\
%4 & 2005-07-17 & 21.8 $\pm$ 0.2 &[-2.17,16.89] & yes?&bckg\\
%  & 2005-10-21 &\\
%  & 2009-07-10 &\\
%5 & 2005-07-17 & $<$ 22.7 $\pm$ 0.3 &[-2.13,16.21] & yes&bckg\\
%  & 2005-10-21 &\\
%  & 2009-07-10 &\\
%6 & 2005-07-17 & $<$ 22.3 $\pm$ 0.3 & [-12.51,-9.25]& yes&bckg\\
  %& 2005-10-21 &\\
%  & 2009-07-10 &\\
%7 & 2005-07-17 & $<$ 22.4 $\pm$ 0.3 & [-7.70,-8.00]& yes&bckg\\
  %& 2005-10-21 &\\
%  & 2009-07-10 &\\

% \enddata
%%\tablecomments{}
%\label{candcomp}
%\end{deluxetable}

\end{document}